\documentclass[%
reprint,
showpacs,
nofootinbib,
 amsmath,amssymb,
 aps,
]{revtex4-1}
\usepackage{graphicx}
\begin{document}

%
%

\title{Stochastic Dynamics and Combinatorial Optimization}

\author{
Igor V. Ovchinnikov}
\author{Kang L. Wang}
\email{wang@seas.ucla.edu}
\affiliation{Electrical Engineering Department,  
            University of California at Los Angeles,
            Los Angeles, CA, 90095}


\begin{abstract}
Natural dynamics is often dominated by sudden nonlinear processes such as neuroavalanches, gamma-ray bursts, solar flares \emph{etc}. that exhibit scale-free statistics much in the spirit of the logarithmic Ritcher scale for earthquake magnitudes. 
On phase diagrams, stochastic dynamical systems (DSs) exhibiting this type of dynamics belong to the finite-width phase (N-phase for brevity) that precedes ordinary chaotic behavior and that is known under such names as noise-induced chaos, self-organized criticality, dynamical complexity \emph{etc.} Within the recently formulated approximation-free supersymemtric theory of stochastics, the N-phase can be roughly interpreted as the noise-induced "overlap" between integrable and chaotic deterministic dynamics. As a result, the N-phase dynamics inherits the properties of the both. Here, we analyze this unique set of properties and conclude that the N-phase DSs must naturally be the most efficient optimizers: on one hand, N-phase DSs have integrable flows with well-defined attractors that can be associated with candidate solutions and, on the other hand, the noise-induced attractor-to-attractor dynamics in the N-phase is effectively chaotic or a-periodic so that a DS must avoid revisiting solutions/attractors thus accelerating the search for the best solution. 
Based on this understanding, we propose a method for stochastic dynamical optimization using the N-phase DSs. This method can be viewed as a hybrid of the simulated and chaotic annealing methods. Our proposition can result in a new generation of hardware devices for efficient solution of various search and/or combinatorial optimization problems.
\end{abstract}

\maketitle

\section{Introduction}
\label{intro}
Many important modern problems including supervised machine learning, Internet and air- traffic control, power grid optimization, \emph{etc}. can be formulated as combinatorial optimization problems (COPs), \emph{i.e.}, the problems of finding the best solution in a large countable set of candidate solutions. The classical example of a COP problem is that of a traveling salesmen. 

In the most general case, when there is no any simplifying structure in the problem, only the brute force method known as the "exhaustive search" can guarantee that the best solution is found: one has to check all the candidate solutions and see which one among them is the best. If the set of candidate solutions is very large, however, the straightforward exhaustive search can be very demanding computationally.

In applications, simplified versions of COP are more practical. One of such versions is to find a good enough solution within a limited time allotted for the search, rather than the best solution ever possible. For simplified versions of COP, various algorithms can be advantageous over the exhaustive search and one of the most successful such algorithms is the family of the annealing techniques including the simulated or thermal annealing (TA) \cite{Kirkpatrick} and chaotic annealing (CA). \cite{Chen_Aihara,Tokuda_Aihara_Nagashima}

From the point of view of the theory of (stochastic) dynamics, the TA and CA algorithms correspond respectively to the stochastic Langevin (purely dissipative) and deterministic chaotic dynamics (see, \emph{e.g.}, Refs.\cite{Chaos_1,Chaos_2} and Refs. therein). These two types of dynamical behavior (stochastic Langevin and deterministic chaotic) are well understood theoretically by now thus providing the TA and CA with firm theoretical foundations. As to the more general types of stochastic dynamics, its use for the optimization purposes has not been investigated in sufficient details yet. One of the reasons for it may be the absence of a reliable theory of stochastic dynamics that would treat non-integrability (or chaos) and noise on equal footing.

Such theory has been worked out very recently (see, \emph{e.g.}, Ref.\cite{Entropy} and Refs. therein) as a result of the conjecture that the mathematical essence of the so-called self-organized criticality \cite{SOC} is the instanton-mediated spontaneous breakdown of the topological supersymmetry that all stochastic differential equations (SDEs) possess.\cite{SOCTFT} This theory is an approximation-free and coordinate-free theory of SDEs that can be dubbed the supersymmetric theory of stochastics (STS) and that has provided a rigorous stochastic generalization of the concept of deterministic dynamical chaos.\cite{Kang} Another important result from the STS,\cite{Book} which is directly relevant to this paper, is revealing the theoretical essence of the stochastic dynamical behavior known as noise-induced chaos,\cite{Gan_Yang_Lei} self-organized criticality,\cite{SOC} intermittency\cite{Pradas,Bashkirtseva_2012} \emph{etc}. and that we call for brevity the N-phase. 

More specifically, all SDEs possess topological supersymmetry and its spontaneous breakdown is the mathematical essence of the dynamical long-range order (DRLO) known under such names as turbulence, chaos,  self-organization {\emph{etc.} As compared to the deterministic case, where dynamical systems (DSs) are chaotic because non-integrable (or chaotic) flow vector fields break topological supersymmetry, in the presence of a (weak) noise there appears the noise-induced chaotic phase, or the N-phase, that precedes ordinary chaos (see Fig.\ref{Figure_1}) and where the topological supersymmetry is broken by the condensation of the noise-induced sudden or (anti-)instantonic tunnelling processes. Examples of such processes include neurodynamical avalanches,\cite{Beggs_2003,Beggs_2004,Chialvo,Brain} earthquakes,\cite{Bak_Earth,Gutenberg} sudden reconfigurations in glasses\cite{Glass} and  biological\cite{Bio} and celestial\cite{Astro} evolutions, solar flares, gamma-ray bursts and other nonlinear processes in the Universe.\cite{XYZ} Since the condensation of these processes is the reason for the DLRO, they often exhibit a scale-free statistics. As a result, the N-phase DSs hosting these processes are often said to be self-organized critical.\cite{SOC}

\begin{figure}
\center{ \includegraphics[width=6cm,height=4cm]{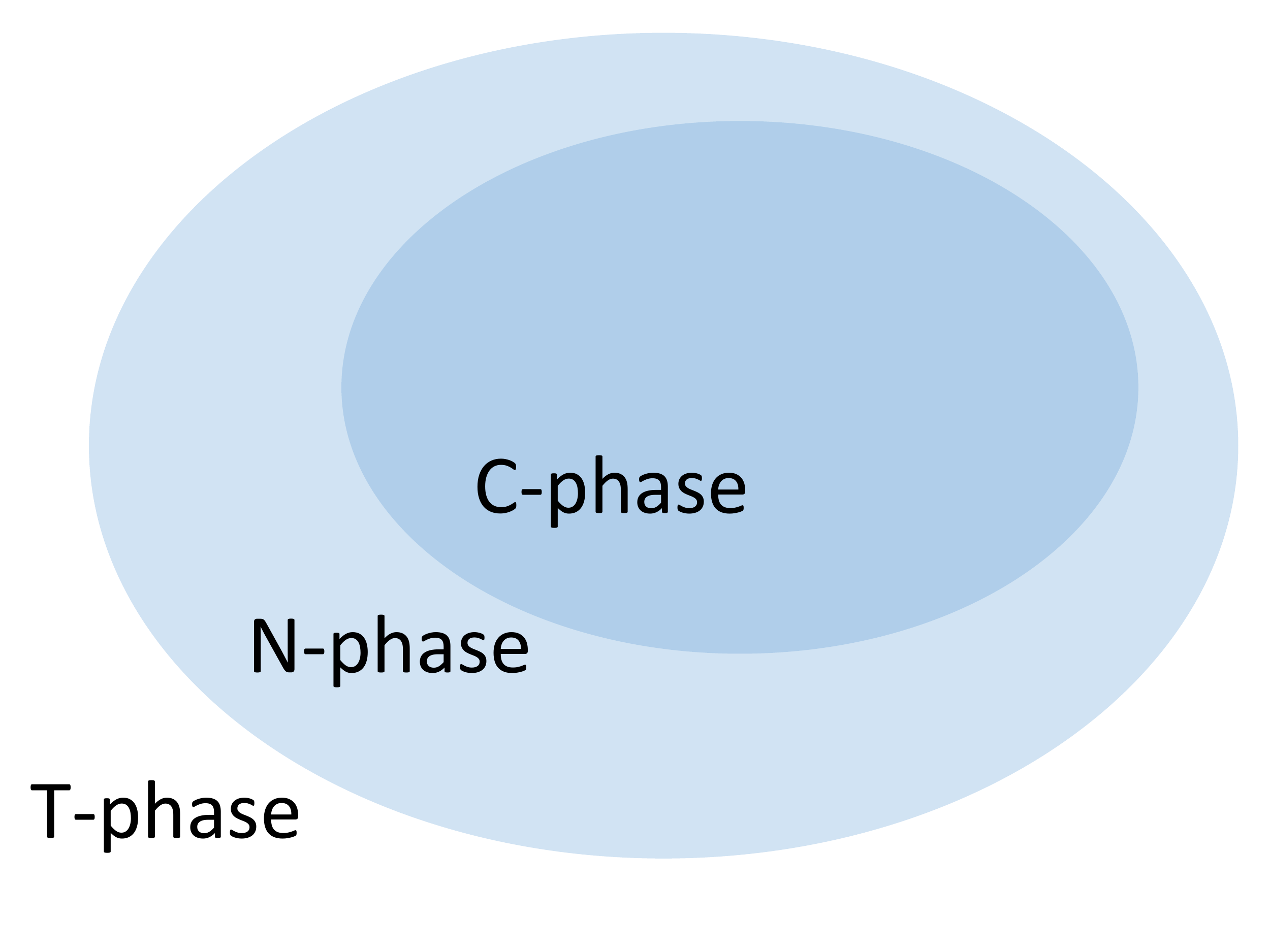}}
\caption{In the weak noise regime, the general phase diagram of stochastic dynamics consists of the three phases: the thermodynamic equilibrium phase (T) with unbroken topological supersymmetry, the N-phase with the topological supersymmetry broken by the noise-induced tunneling processes, and the ordinary chaotic phase (C) with the topological supersymmetry broken by the non-integrability of the flow vector field. In the deterministic limit, the N-phase collapses/shrinks into the boundary between the T- and C- phases.}
\label{Figure_1}
\end{figure}

Just as many other DSs in the N-phase, biological evolution\cite{Bak_Evolution} may be expected to be an efficient optimizer. In this line of thinking, the bio-inspired numerical optimization heuristics called extremal optimization was proposed.\cite{Extremal_Optimization} While the efficiency of the extremal optimization\cite{Demostration1,Demostration2} as well as optimization algorithms mimicking biological evolution more accurately\cite{SOC_EO} is well-established, no theoretical explanation why N-phase dynamics may be advantageous for the optimization purposes existed previously. In this paper, we use the STS picture of the N-phase dynamics to cover this gap.

The simplified version of the STS picture of the N-phase dynamics is a noise-induced overlap between chaotic (non-integrable) and regular (integrable) dynamical behaviors. As a result, this phase inherits the properties of the both. Our analysis will point onto the possibility that this unique set of properties includes those that are advantageous for the purpose of the search of solutions and is free from those that can be viewed as drawbacks. In other words, DSs in the N-phase are naturally the best optimizers. Based on this understanding, we propose a method that can be called the N-phase annealing (NA) and/or searching. The NA can be identified as a hybrid of the TA and CA methods mentioned above. We believe that this idea can form a core for the new generation of the efficient hardware optimizers.

The rest of the paper is organized as follows. In Sec. \ref{sec:Gen_Phase_Diag}, the general phase diagram of stochastic DSs from the STS is briefly outlined. In Sec. \ref{sec:Annealing}, the STS is used to analyze the TA and CA methods and qualitatively discuss their advantages and drawbacks for optimization. In Sec. \ref{sec:N_Annealer}, the possibility of using the N-phase DSs for optimization is considered and it is argued that this approach has the advantages of both the TA and CA methods and is free of their drawbacks. Sec. \ref{sec:Conclusion} concludes the paper. 

\section{General Phase Diagram of Stochastic Dynamics}
\label{sec:Gen_Phase_Diag}
\subsection{Ubiquitous Supersymmetry and its Spontaneous Breakdown}

The recently found approximation-free STS\cite{Entropy} lays at the heart of the proposition in this paper and we would like to begin with a brief discussion of this theory. The easiest way to address the STS is to consider the following class of SDEs that covers most of the models in the literature,
\begin{eqnarray}
\dot x(t) = {F}(x(t)) + (2\Theta)^{1/2}e_a(x(t)) \xi^a(t). \label{SDE}
\end{eqnarray}
Here and in the following the summation is assumed over the repeated indices; $x\in X$ is a point in the phase space, $X$, which is assumed to be a topological manifold; $F({x})$ is a flow vector field on $X$ at $x$; ${e}_a({x}), a=1,...$ is a set of vector fields on $X$ with $a$ being the parameter running over them; $\xi^a(t)$ is a set of the Gaussian white noise variables with the standard stochastic expectation values $\langle \xi^a(\tau_1)\rangle = 0$ and $\langle \xi^{a_1}(\tau_1)\xi^{a_2}(\tau_2)\rangle = \delta^{a_1a_2}\delta(\tau_1-\tau_2)$.

As discussed in details in the Appendix, Eq.(\ref{SDE}) uniquely defines the temporal evolution of differential forms, $\psi$'s, on $X$, regarded as wavefunctions,
\begin{eqnarray}
\partial_t \psi = - \hat H \psi,
\end{eqnarray}
where the (infinitesemal) stochastic evolution operator (SEO), $\hat H$, is defined in Eq.(\ref{SEO}). The SEO intrinsically possesses the topological supersymmetry represented by de Rahm operator known also as exterior derivative, $\hat d$. As a result, all eigenstates are divided into two categories: supersymmetric siglets and non-supersymmetric doublets. All supersymmetric singlets have zero eigenvalue, whereas the ground state(s) of the model is the one(s) with the least real part of its eigenvalue. Thus, when the SEO spectrum has one of the two forms given in the middle and bottom graphs of Fig.\ref{Figure_2}a, the ground state(s) is non-supersymmetric and it is said that the $d-$symmetry is broken spontaneously. Note that both presented types of the spontaneous $\hat d$-symmetry breaking spectra are realizable as was recently established in Refs.\cite{Torsten,Sun}.

The existence of this $\hat d$-symmetry in all SDEs is merely the algebraic representation of the "phase-space continuity" of the SDE-defined dynamics. In other words, for any configuration of the noise, two infinitely close points in $X$ will remain close. When the $\hat d$-symmetry is broken spontaneously, this property is violated in the limit of the infinitely long dynamical evolution, represented, of course, by a non-supersymmetric ground state. In other words, for DS with the spontaneously broken $\hat d$-symmetry, the DS exhibits the butterfly effect because two close points may result is very different trajectories in the limit of the infinitely long evolution. Therefore, the phenomenon of the spontaneous breakdown of the $\hat d$-symmetry, or the DRLO, is the stochastic generalization of the concept of deterministic chaos.\cite{Kang} 

The importance of the $\hat d$-symmetry breaking picture of dynamical chaos is twofold. First, unlike the classical trajectory-based picture, the $\hat d$-symmetry breaking picture works just as well for stochastic cases. Second, the $\hat d$-symmetry breaking picture immediately suggests that due to the Goldstone theorem DSs with broken supersymmetry must exhibit the long-range dynamical behavior (LRDB) in that or another form including the butterfly effect or the sensitivity to initial conditions and/or perturbations, 1/f noise or the long-term memory effect, or the scale-free statistics of avalanche-type sudden processes. Therefore, say, the butterfly effect is essentially a consequence of the DRLO rather than the part of its definition, as is often pictured in the literature.

Yet another important insight provided by the $\hat d$-symmetry breaking picture of DRLO is the following. In physics, the low-symmetry phases with spontaneously broken symmetries possess "order" as compared to the high-symmetry phases with unbroken symmetries. An example from statistical physics is the ferromagnetic phase with the spontaneously broken $SU(2)$ symmetry of the global rotations of spins. As compared to the paramagnetic phase, ferromagnets have magnetic order, or ferromagnetism, that can be described by a ferromagnetic order parameter -- the local magnetization of spins. 

Similarly, it is actually the chaotic phase, which is the low-symmetry or ordered phase as compared to the phase of thermodynamic equilibrium, \emph{i.e.}, the phase with unbroken $\hat d$-symmetry. From this perspective, (stochastic) chaos may not be the best name for this dynamical phenomenon as "chaos" literally means "the absence of order". \footnote{This 'ordered' picture of chaos goes well with the topological theory of deterministic chaos\cite{Gilmore} studying non-trivial topological structures (links, knots \emph{etc}.) produced by unstable periodic orbits of deterministic chaotic flows.} This is the reason why we use DLRO instead of another commonly used identifier of stochastic chaos.

\subsection{The Phase Diagram}
\label{sec:Phase_Diagram}

The STS allows for one step further in the direction of understanding qualitatively different types of stochastic dynamics. A refined classification of stochastic DSs beyond the broken/unbroken $\hat d$-symmetry follows from the analysis of the possible ways in which the $\hat d$-symmetry can be spontaneously broken (Fig.\ref{Figure_2}b). 

In the deterministic limit, the classification is trivial: the flow vector field is either integrable or chaotic (non-integrable). In the presence of noise, however, there appears yet another mechanism of $d$-symmetry breaking. This mechanism is the noise-induced tunneling processes between, say, different attractors. This mechanism is known in the high-energy physics as the condensation of (anti-)instantons, which is a technical term for the tunneling processes between different “perturbative” vacua representing, say, different attractors of the flow. Note also that supersymmetries are hard to break perturbatively due to what is known as the non-renormalization or no-anomaly theorems.\cite{Seiberg} Thus, the (anti-)instanton condensation is, perhaps, the most reliable mechanism for the spontaneous breakdown of a supersymmetry in high-energy physics models.\cite{Witten} 

The phase with $\hat d$-symmetry spontaneously broken by the noise-induced tunneling processes is what we call the N-phase. For an external observer, the dynamics in the N-phase looks like a sequence of unpredictable tunneling events and/or avalanches, with neuroavalacnhes in brain being one example of the N-phase dynamics. The position of the N-phase right before the ordinary chaotic phase (C-phase) is not accidental. Indeed, in the deterministic limit, when the noise-induced tunneling disappears, the N-phase must collapse onto the boundary between the phase of thermodynamic equilibrium (T-phase) and the C-phase (see Fig. \ref{Figure_2}b).

\begin{figure}
\center{ \includegraphics[width=8.6cm,height=5.5cm]{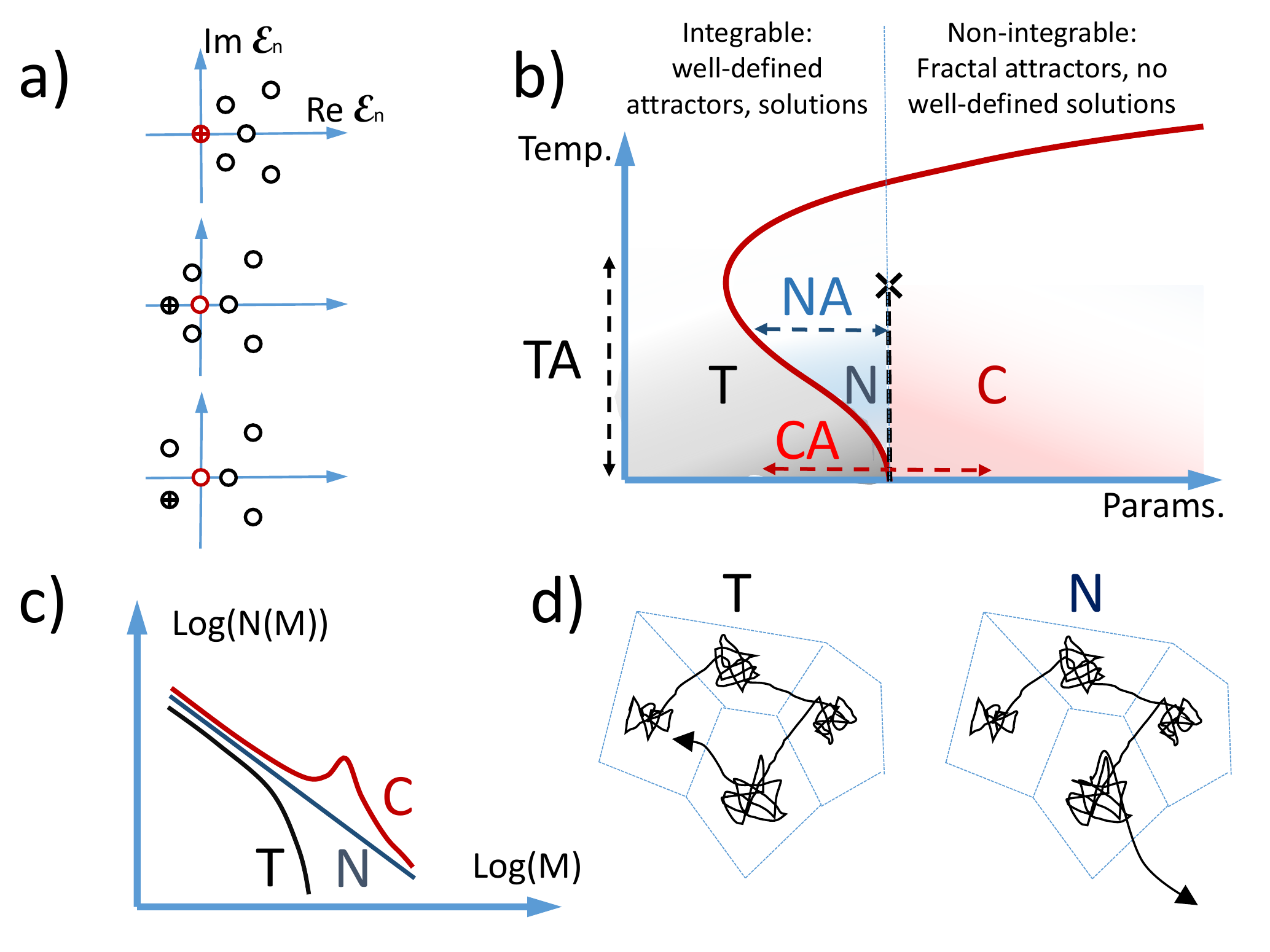}}
\caption{{\bf (a)} The three types of spectra of stochastic evolution operator: with unbroken (top, thermodynamic equilibrium) and spontaneously broken (middle, and bottom) $\hat d$-symmetry. The hollow circles represent eigenstates and (red) circles at the origin represent the supersymmetric eigenstates. The circles with the crosses inside represent the ground states of the model. {\bf (b)} General phase diagram of a stochastic DS on the temperature-parameters space. Thermodynamic equilibrium (T) phase has unbroken $\hat d$-symmetry; the phase of the “ordinary” chaos (C) has $\hat d$-symmetry broken by chaotic or non-integrable flow; the N-phase has the $\hat d$-symmetry broken by the condensation of noise-induced tunneling processes (neuroavalanches in case of brain). In deterministic limit, the N-phase collapses into the boundary between the C- and T- phases.  At higher temperatures, the sharp boundary between the N- and C-phases smears out into a crossover. The thermal annealing (TA) and chaotic annealing (CA) methods correspond respectively to the high-temperature T- phase and the zero temperature C-phase.  The single-run regime is realized as the “in-and-out” manipulation of the parameters (double-arrows). The N-phase phase annealing (NA) method has both features of the well-defined attractors of the TA and the "chaotic" long-range order of the CA. {\bf (c)} The three low-temperature phases exhibit different statistics of (the masses of) avalanches (M): N-phase, a power law; C- and T-phases, a bump and a cut-off on the large-avalanche side, respectively. {\bf (d)} Major difference between dynamics in the T-phase (left) and the N-phase (right) is that the dynamical or chaotic long-range order of the N-phase reveals itself in non-revisiting the attractors. This must speed up the solution search for the NA as compared to the TA method. At the same time, unlike the CA method, the NA has well-defined attractors that can be associated with candidate solutions.}
\label{Figure_2}       
\end{figure}

One important piece of understanding is that because the $d$-symmetry is spontaneously broken in the N-phase by the tunneling processes, these processes must exhibit signatures of the emergent DLRO. One of these signatures is the well-established scale-free or power-law statistics of various characteristics of the tunneling processes such as their "masses" (Fig.\ref{Figure_2}c).

A less important point for our purposes is the opposite limit of high temperatures. In this limit, the “perturbative” ground states on different attractors of the integrable flow-vector field overlap too much so that an external observer will not be able to tell one tunneling process from another. As a result, the sharp boundary between the N- and C-phases must smear out into a crossover and the two phases must merge into a single phase that can probably be identified as the strong-coupling regime of the DLRO.  
Finally, at even higher temperatures, the $\hat d$-symmetry must eventually be restored because the dynamics will be dominated by diffusion, which alone should not break the $\hat d$-symmetry as physically sound diffusion must correspond to Laplacians with real and non-negative spectrum that does not break $\hat d$-symmetry spontaneously.

\section{Chaotic and Thermal Annealing Methods}
\label{sec:Annealing}

In this Section, we would like to discuss the TA and CA techniques from the point of view of STS. This will allow later to analyze the advantages and drawbacks of these two methods and address the question of the feasibility of the implementation of the annealing technique using the N-phase DSs.

\subsection{Thermal Annealing: Thermodynamic Equilibrium}

TA is the most celebrated method of optimization in the literature. For example, the famous D-wave machine,\cite{DWave} that was originally designed for quantum computing, has been utilized so far only as an implementation of the TA method, though with an additional quantum tunneling mechanism. 

The essence of the TA\cite{Kirkpatrick} is the numerical or experimental realization of a Langevin SDE. This class of SDE is featured by flow vector fields that are gradients of some function called Langevin function. In most of the examples in the literature, the phase space is linear, $X=\mathbf{R}^N$, the metric is Euclidian so that $e^i_a = \delta^i_a$ in Eq.(\ref{SDE}) and the flow vector field is, $F = - \delta^{ij} \partial U(x)/\partial x^j$, where the Langevin function, $U$ , is highly nonlinear with multiple local minima that can be associated with the set of candidate solutions of the COP at hand.

From the point of view of the STS, the $\hat d$-symmetry is never broken for physically meaningful Langevin SDEs.\cite{Entropy} The TE state is (one of) the ground state(s) of the DS with the Langevin potential playing the role of the Lyapunov function,
\begin{eqnarray}
\psi_{TE} \propto e^{-U(X)/\Theta}dV,
\end{eqnarray}
where $dV = dx^1\wedge...\wedge dx^{\mathrm{Dim }X}$ is the volume element. 

The common operation of the TA algorithm can be described as a “single run” method - each run provides one candidate solution.  The procedure is as follows (see Fig.\ref{Figure_2}b, the arrow in and out of the high temperature regime). One brings the DSs to the state of sufficiently high temperature. After the DS has thermalized to its TE state, the temperature is gradually reduced to zero. At that, the TE state is being adiabatically changed into a $\delta$-function-like distribution, $\psi_{TE}\sim \delta(x-x_{g-m})dV$, where $x_{G-M}$ is the position of the global minimum of $U(x)$.  The reduction of the temperature must be sufficiently slow in order to assure that the DS adiabatically remains in its TE state during the cooling process.  The slower the cooling is, the likelier it is that the DS will end up in the true global minimum and not in one of its multiple local minima.

\subsection{Chaotic Annealing: Deterministic Chaos}

As is suggested by its name, the CA method\cite{Chen_Aihara,Tokuda_Aihara_Nagashima} is based on the properties of chaotic dynamics. Unlike the TA method, the CA method is deterministic.  To be more specific, one considers a deterministic DS ($\Theta\to0$ ) with a one-parameter flow vector field in Eq.(\ref{SDE}),
\begin{eqnarray}
F_{\eta}(x) = \eta F_L(x) + (1-\eta) F_C(x), 
\end{eqnarray}
where $F_L^i=-\delta^{ij}\partial U(x)/\partial x^j$ is a dissipative or Langevin part as in the TA method, $F_C$ is some chaotic flow vector field, and $\eta$ is the (homotopy) parameter connecting $F_L$ and $F_C$. Clearly, there is at least one $\eta=\eta_C, 0<\eta_C<1$, at which the transition from integrable to chaotic dynamics occurs.

The idea now is to employ the intrinsic itinerancy of chaotic dynamics for the search of the best solution among all the local minima of the Langevin potential, $U$.  One can try to implement the single run methodology of the TA algorithm as follows (see Fig.\ref{Figure_2}b, the arrow in and out of the deterministic chaos): one moves the DS into the chaotic regime at some, $\eta>\eta_C$, allows the DS to chaoticaly move for some long enough time, and then gradually moves the DS out of the C-phase into the T-phase, where the DS settles into one of the local minima of $U$.  Now, it may seem that the so-found solution is a good candidate for the global minimum of $U$. 

This is not quite correct, however. From the solutions search point of view, the chaoticity of the CA method and the external noise of the TA method may look alike in making the DS itinerant in the phase space so that it could visit many candidate solutions/regions. On the over hand, the solution provided by a single run TA procedure is almost independent of the initial conditions and of how one manipulates the temperature parameter (provided it is sufficiently slow). In other words, the TE state is memoryless. It forgets where the DS began its search and how one manipulated the temperature parameter.  On the contrary, the chaotic dynamics of the CA has the ultimate infinitely long memory of the initial conditions and/or perturbations (the butterfly effect).  This infinite memory is opposed to the memoryless TE.  The solution provided by a single run CA procedure is highly dependent not only on the initial condition, but also on the way one manipulates parameter, $\eta$.  Clearly, a solution that depends strongly on the initial condition and on the behavior of the "driver" cannot be reliable.

As we discuss in Sec.\ref{sec:N_Annealer} below, the chaotic sensitivity to initial conditions that makes the single-run CA unreliable, actually becomes an advantage in the continuous-time search regime, though for a very different reason.

\subsection{Continuous-time search regime of TA}
\label{sec:cont_search}

As we discussed in the previous section, the CA method can not be used in the "single run" regime of the TA method. It is probably for this reason that the CA method is often discussed in the context of the “continuous-time search regime”: it is said that chaotic DS "visits" good candidate solutions during its evolution.\cite{Chen_Aihara,Tokuda_Aihara_Nagashima} Thus, it is only in the continuous-time search regime that the CA and TA methods can be compared. Furthermore, the continuous time search is what we are aiming at in this paper for the N-phase annealing method. Thus, before we can compare the TA and CA methods, the TA method must be adopted for the continuous-time search implementation. 

This can be done as follows. Instead of moving the DS in and out of the high temperature regime, one can fix the temperature at magnitude, which is sufficiently high to allow for tunneling between different local minima. The dynamics will look like a sequence of jumps between the local minima. The DS will essentially serve as a generator of candidate solutions, which may be read out in real time. This sequence can be monitored. When a better solution has been generated, it is updated continuously until a good enough solution has been achieved.

\subsection{Comparison of TA and CA}

An obvious advantage of the TA method is the existence of the set of local minima that can be directly associated with the set of candidate solutions of the COP. This is in contrast with CA method, in which the DSs does not have local minima or, more generally, well-defined attractors that could be associated with the set of solutions.   

On the other hand, chaotic dynamics has its own advantages. First, "chaotic" is antonym for "periodic" as chaotic trajectory never crosses itself. This property is one qualitative way to understand the essence of chaotic dynamics: the DS “remembers” all the points it has previously been to and avoids revisiting them. The TA method, on the other hand, does not possess the memory of its past. In other words, the generation of the sequence of solutions in the TA is essentially “diffusive” in the phase space.  As a result, the dynamics of the DS will not make effort to avoid solutions that have been already visited/generated.

Another valuable property of chaos is due to the so-called "topological mixing", \emph{i.e.}, the property of deterministic chaotic dynamics that any two regions in the phase space are connected by a trajectory. This implies that the phase space region corresponding to the best solution of the COP will be eventually found by the chaotic search no matter where the search has begun. 

The TA method does not have this property. Unlike in the chaotic search, the noise-induced Langevin search is carried out not in a "directional" manner but rather in a diffusive, randomly directed way. As a result, it may take longer for the TA to generate a good enough solution. 

Thus, TA and CA methods both have advantages and disadvantages for the purpose of phase space search. It would be desirable to have a method, which has advantages of the both methods and free of their disadvantages. As we discuss in the next section, the properties of the N-phase DSs that follow from its STS picture do allow for the realization of such an annealing method.

\section{N-Phase Dynamical Search}
\label{sec:N_Annealer}

The N-phase dynamics has been studied in different areas of science mostly numerically and a consistent theory of the N-phase dynamics does exist yet. Such a consistent theory can be achieved within the STS and in the combination with the methodology of the low-energy effective theory. The formulation of the methodology for the construction of the effective low-energy theory of N-phase dynamics, which is, of course, model specific, is actually one of the next important steps in further development of the STS. This work has not been done yet.

Fortunately, we do not need a quantitative theory for N-phase dynamics for the purpose of this paper. The qualitative understanding of the essence of the N-phase dynamics discussed in Sec.\ref{sec:Phase_Diagram} is already enough to see that the dynamical optimization using N-phase DSs inherits the advantages of the both the TA and the CA methods and not their drawbacks. More specifically, on the one hand, just like the DSs in the T-phase, but unlike those in the C-phase, the DSs in the N-phase have well-defined attractors because the flow vector field is integrable. These attractors can be directly associated with the set of candidate solutions of the COP. On the other hand, unlike the DSs at the T-phase but just like the DSs at the C-phase, the N-phase DSs have their $\hat d$-symmetry spontaneously broken so that they must possess the properties of deterministic chaotic behavior and the advantages associated with it as discussed in the previous section. Namely, the property of the chaotic “non-periodicity” must, most likely, transform into the tendency of the DSs to avoid revisiting attractors/solutions as schematically demonstrated in Fig.\ref{Figure_2}d. This must accelerate the search for a good solution.

The above reasoning is a very general and works without the specification of the COP at hand. The next question is how to harness the advantages of the N-phase dynamics for a specific COP. We believe that this can be achieved with the help of the feedback control of the N-phase dynamics as discussed in the rest of this Section.  

\subsection{N-Phase Dynamical Optimization Unit}
\label{sec:N-Annealer_Unit}

Based on the above understanding, we conceive an implementation of the N-phase dynamical optimization as illustrated in Fig. \ref{Figure_3}a. Let us refer to this implementation to as the N-phase Optimization Unit (NOU).  An NOU consists of the two major parts: the N-phase Dynamical Solution Generator (N-DSG) and the software part. The N-DSG is the heart of the device. It is a physical DS in its N-phase. We will speak of the functionality of N-DSG shortly. Let us first discuss briefly one of its possible physical implementations.

Clearly, there are abundant nanoscale engineered DSs, which can be used as an N-DSG. Examples of the N-DSGs may include those of the coupled oscillators which may be fabricated from CMOS circuits, spin torque oscillators and other nanoscale dynamic elements. For the sake of concreteness, we think of our N-DSG as of a lattice of the carefully designed coupled ferromagnetic nano-disks brought into its N-phase by an externally controlled electric current.

One type of the N-phase dynamics of an N-DSG is a sequence of “patterns” or matrices of zeros and ones realized as ups and down spins in a ferromagnetic lattice. The patterns change suddenly through avalanches that can be characterized by their masses, \emph{i.e.}, by the number of units switched at this avalanche. The location of the N-phase on the space of the externally controlled parameters, \emph{e.g.}, the magnitude of the driving electric current, must be established at the calibration level using the scale-free avalanche statistics criterion illustrated in Fig.\ref{Figure_2}c.

This sequence of patterns is read out by the second module of the NOU -- the software part of the device. In case when the N-DSG is implemented as an array of magnetic elements, the read out can be experimentally realized via the magneto-optic Kerr effect (MOKE) microscope, which transfers magnetic patterns into an optical image of ups and downs.

\subsection{Patterns as Solutions}

The software part of the NOU interprets patterns that are sequentially arriving from the N-DSG as candidate solutions for the COP at hand, which in its turn, is implemented on the software level. Here is an example how 2D patterns or rather matrices of zeros and ones can be interpreted as candidate solutions of a traveling salesman problem (TSP). The TSP asks the following question: given a set of cities and the pair-wise distances between them, what is the closed path that visits each city exactly once and has the least possible distance? There are $(N-1)!$ different closed paths obtained by all possible permutations of cities, $p$'s, in the closed path $123…(N-1)$, with numbers being the indices of the cities. The cost function for each path $p(12…(N-1))$ is the total distance traveled: $C(p)=\sum_{i=1}^{N-1}d_{p(i)p(i+1)}$, where $d$’s are the pair-wise distances between the cities. Now, for each pattern from the N-DSG, or for any matrix of zeros and ones, one can construct a set of $N-1$ numbers that are the sums of the columns (or rows) of this matrix, $M_{ij}\to R_i=\sum_{ij}M_{ij}$. The candidate solution to the TSP is the permutation that renders them in ascending order, $p(R_i): R_{p(1)}<...<R_{p(N-1)}$. The cost function, $C(p)$, is then minimized.

In fact, the correspondence law between 2D patterns and a finite countable set of candidate solutions can be devised for any other COP. After all, a 2D pattern is just a (long) binary number and any set of objects like candidate solutions for a given COP can be embedded into the space of all possible patterns, provided that the number of elements in the N-SDG is large enough. Such pattern-solution correspondence is, of cause, COP-specific. 

\begin{figure}
\center{ \includegraphics[width=8.6cm,height=6.5cm]{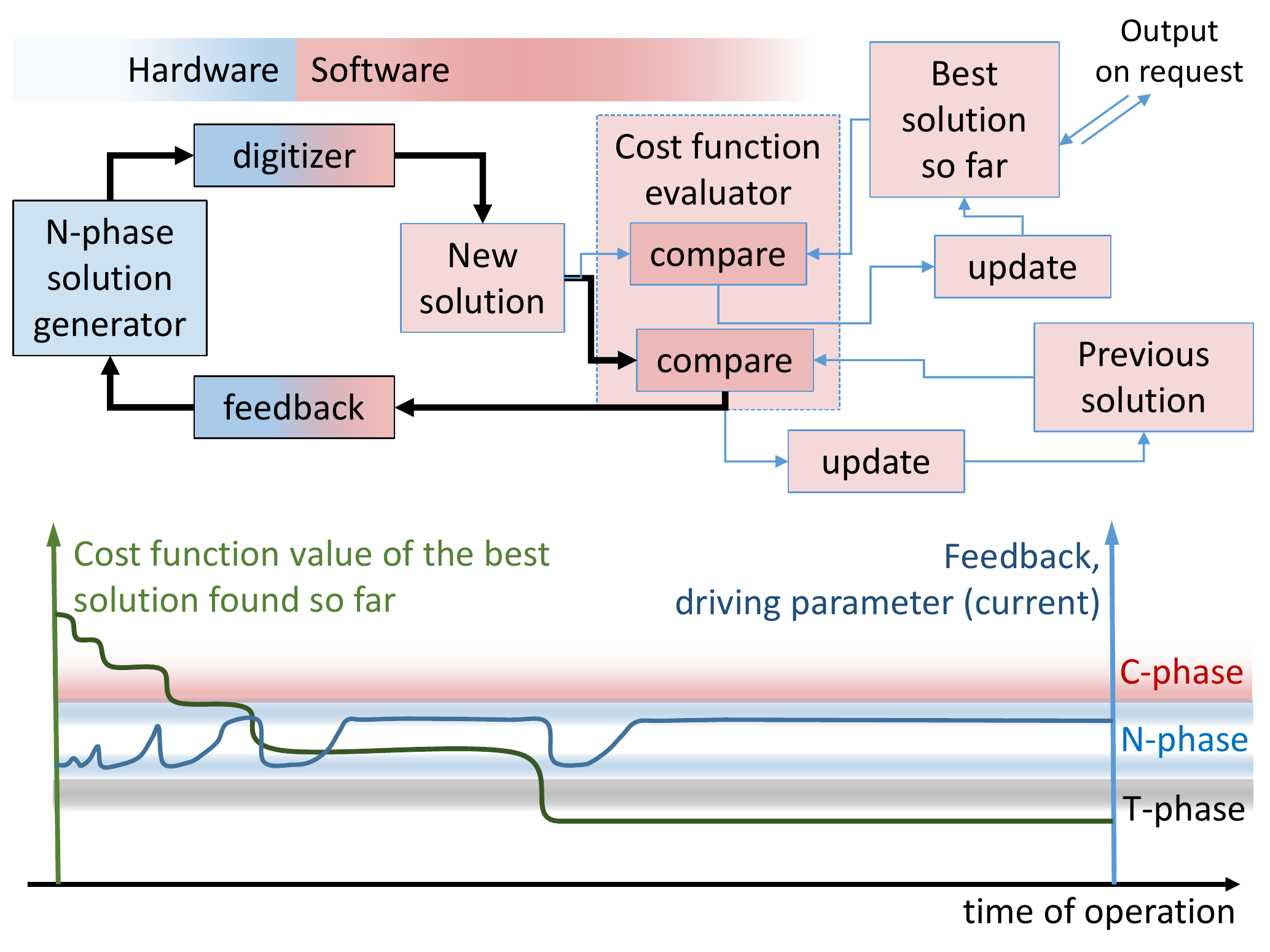}}
\caption{{\bf (Upper graph)} The heart of the NOU is the N-phase dynamical solution generator (N-DSG), which continuously generates candidate solutions for the combinatorial optimization problem at hand, implemented as the cost function in the software part of the device. The output of the device is the best solution found so far. b. {\bf (Lower graph)} Depending on either the new solution is better/worse than the previous solution, the software part shifts the control parameter of the N-DSG toward the C-phase or T-phase, respectively.  The control parameter makes the solution generator more or less chaotic depending on either the new solution is better or worse than the previous solution. The goal of this feedback control is to allow the N-DSG to explore the phase space (nearby solutions) more carefully if the new solution is an improvement. On the contrary, if the new solution is a worsening, the control parameter makes the N-DSG travel faster through this region of the phase space.}
\label{Figure_3}
\end{figure}

\subsection{N-phase Operation of NOU}

Now, after having discussed the relation between the patterns of the N-DSG and candidate solution of the COP at hand, one can proceed with the discussion of the dynamical operation of the NOU (Fig.\ref{Figure_3}b). Once a new candidate solution is generated by N-DSG and received by the software, the cost function of the new solution is evaluated. The software part compares the new cost function with that of the previous solution and that of the best solution found so far, which is updated accordingly if the new solution turns out to be better. Depending on whether the new solution is better or worse than the previous solution, the software part shifts the control parameter of N-DSG (\emph{e.g.}, the driving electrical current) toward the C-phase or the T-phase, respectively. The control parameter makes the N-DSG more or less "chaotic" allowing the solution generator to explore the phase space, or rather nearby solutions, more thoroughly in case the new solution is an improvement. On the contrary, if the new solution is worse, the control parameter forces N-DSG to pass this part of the phase space faster.

The N-DSG can in principle be substituted by a T-phase stochastic DS, which would act as a TA optimizer operating in the continuous-time regime as discussed at the end of Sec.\ref{sec:cont_search}. The use of the T-phase DS will certainly worsen the performance. A quantitative estimation of the improvement of the N-phase optimization over the T-phase optimization is not available at this moment. Nevertheless, it is reasonable to expect that this improvement must be of the same significance as that in the three-coloring graph problem with the extremal optimization procedure by turning to a power-law distribution of individual cells.\cite{Optim} There, the local minima of the problem can be efficiently sampled in $O(n^{3.5})$ as compared to the exponential growth in the non-extremal optimization scheme. In our case, the N-phase with the DLRO must be paralleled with the extremal (or power-law) optimization in the coloring problem above. Therefore, it is reasonably to expect a significant improvement due to the use of a N-phase DS.

\section{Conclusion}
\label{sec:Conclusion}

The recently proposed approximation-free supersymmetric theory of stochastics has revealed that on the border of “ordinary” chaotic dynamics there exist a full-dimensional dynamical phase, that we call the N-phase, with a unique set of dynamical properties. In this paper, we argued that this set of properties naturally makes the DSs at the N-phase the most efficient optimizers or search engines. Based on this understanding, we proposed the dynamical optimization method using the N-phase DSs, which can be viewed as a hybrid of the chaotic and simulated annealing optimization methods. This method must inherit the advantages and not the drawbacks of the two parental annealing techniques. In particular, the N-phase DSs possess dynamical long-range order just like ordinary chaotic DS, which must result in the acceleration of the best solution search. On the other hand,  just like in the thermal annealing method, the flow vector field is integrable so that the N-phase DSs have well defined attractors that can be directly associated with the set of candidate solutions of the optimization problem at hand.

We would also like to mention here yet another probably less concrete but not any less important motivation of this paper. The part of the physics community studying stochastic dynamics in brain has already established that brain is a DS in the "self-organized critical" phase, or N-phase in our terms. One natural question that immediately arises is whether it is possible to identify the properties of the N-phase dynamics that make it preferable for nature to make brain a N-phase DS. To the best of our knowledge, a convincing answer to this question has not been found yet. Our proposition that the N-phase DSs are the best dynamical optimizers may, perhaps, be looked upon as a first attempt to answer this question from the mathematical or physical point of view.

\section*{Acknowledgments}
KLW would like to acknowledge the endowed support of Raytheon Chair Professorship.

\appendix

\section{STS in a nutshell}

\subsection{Stochastic Evolution Operator}
The stochastic evolution operator (SEO) is the centerpiece of the STS. It can be obtained in the following four steps (see, \emph{e.g.}, Ref. \cite{Entropy} for details):

$1)$ For a fixed configuration of noise, Eq.(\ref{SDE}) is a time-dependent ordinary differential equation that uniquely defines a trajectory, $x(t)$, given an initial condition, $x(t)|_{t=t'}=x_0$. 

$2)$ These trajectories can be looked upon as a two-parameter family of diffeomorphisms of the phase space on itself, $M_{tt'}:X\to X$. The above trajectories can be given as $x(t) = M_{tt'}(x_0)$.

$3)$ The above SDE-defined diffeomorphisms induce actions or pullbacks on the elements of the exterior algebra of the phase space, which is considered a Hilbert space of the model,
\begin{eqnarray}
M^{*}_{t't}:\Omega\to\Omega,
\end{eqnarray}
where $\Omega= \bigoplus_{k=0}^{D}\Omega^k$ with $\Omega^k$ being the space of differential forms of degree $k$ or $k$-forms of various degrees, 
\begin{eqnarray}
\psi^{(k)} = \psi^{(k)}_{i_1...i_k}dx^{i_1}\wedge...\wedge dx^{i_k}\in \Omega^{k}.\label{k_form}
\end{eqnarray}

$4)$ Unlike trajectories in the general case of a nonlinear phase space, the pullbacks, together with $\Omega$, are always linear objects. As such, one can perform stochastic averaging over the configurations of the noise and arrive at the finite-time SEO:
\begin{eqnarray}
\hat{\mathcal{M}}_{tt'}: \Omega &\to& \Omega, \hat{\mathcal{M}}_{tt'} = \langle M^*_{t't}\rangle.
\end{eqnarray}
A bit lengthy but straightforward calculation leads to the finite-time SEO in the following form,
\begin{eqnarray}
\hat{\mathcal{M}}_{tt'} = e^{-(t-t')\hat H},
\end{eqnarray}
where the (infinitesemal) SEO can be given as,
\begin{eqnarray}
\hat H = \hat {\mathcal L}_{F} - \Theta \hat {\mathcal L}_{e_a} \hat {\mathcal L}_{e_a} = [\hat d, \hat{\bar d}],\label{SEO}
\end{eqnarray}
where $\hat {\mathcal L}_F$ is the Lie derivative along $F$; $\hat {\bar d} = \hat\imath_F - \Theta \hat \imath_{e_a}\hat{\mathcal L}_{e_a}$ is the operator that can be identified as the probabaility current operator with $\hat\imath_F \equiv F^i \hat\imath_i$ and 
\begin{eqnarray}
\hat\imath_i:\Omega^{k}\to\Omega^{k-1}, \hat\imath_i\psi^{(k)} = k \psi^{(k)}_{i i_2...i_k}dx^{i_2}\wedge...\wedge dx^{i_k},
\end{eqnarray}
being the operator of the interior multiplicaiton, and 
\begin{eqnarray}
\hat d: \Omega^k\to\Omega^{k+1}, \hat d \psi^{(k)} = \frac{\partial\psi^{(k)}_{i_1...i_k}}{\partial x^i}dx^i\wedge dx^{i_2}\wedge ...\wedge dx^{i_k},
\end{eqnarray}
is the exterior derivative or de Rahm operator. In the two above formulas, $\psi^{(k)}$ is given in  Eq.(\ref{k_form}) and Cartan formula has been used,
\begin{eqnarray}
\hat {\mathcal L}_F = [\hat d, \hat\imath_F].\label{Cartan}
\end{eqnarray}
The notation for the commutator above denotes the bi-graded commutator, which is an anticommutator if both operators are fermionic, \emph{i.e.}, of odd degree as in Eq.(\ref{SEO}), and it is the commutator otherwise. This concludes the derivation of the SEO of SDE (\ref{SEO}).

\subsection{Spontaneous Breakdown of Supersymmetry}

The exterior derivative is a symmetry or, rather, a supersymmetry of all the SDEs. This is seen from the commutativity of the SEO with the exterior derivative,
\begin{eqnarray}
[\hat H, \hat d] = 0.
\end{eqnarray}

In physics, symmetries of evolution operators reveal themselves via protected degeneracies of the eigenstates, \emph{i.e.}, eigenstates of degenerate eigenvalues form multiplets that are irreducible representations of the corresponding symmetry group. Accordingly, there are two types of eigenstates of the SEO. Some of the eigestates are supersymmetric singlets, $|\theta\rangle$'s, that are non-trivial in de Rahm cohomology, \emph{i.e.}, $\hat d|\theta_i\rangle=0$ and no such state  $|\theta_i'\rangle$ exists that $|\theta_i\rangle=\hat d|\theta_i'\rangle$. Each de Rahm cohomology class provides one supersymmetric eigenstate and all of them have exactly zero eigenvalue. 

All the other eigenstates, of which there are infinite number, of course, are the non-supersymmetric doublets of the form $|\vartheta\rangle$ and $\hat d|\vartheta\rangle$. If the ground state of the SEO, \emph{i.e.}, the eigenstate with the lowest real part of its eigenvalue, is a non-supersymmetric doublet as in the middle and bottom graphs of Fig.\ref{Figure_2}a, the $\hat d$-symmetry is said to be broken spontaneously. It can be shown that such situation can be looked upon as the stochastic generalization of the concept of deterministic chaos.\cite{Kang}

\end{document}